\documentclass[a4paper,11pt]{article}
\usepackage{jcappub}

\usepackage[colorlinks=true,urlcolor=blue,linkcolor=blue]{hyperref}
\usepackage{graphicx}
\usepackage{epstopdf}
\usepackage{color}
\usepackage{epsfig}
\relax

\def\be{\begin{equation}}
\def\ee{\end{equation}}
\def\bs{\begin{subequations}}
\def\es{\end{subequations}}

\def\Lx{\Lambda}

\def\be{\begin{equation}}
\def\ee{\end{equation}}
\def\bs{\begin{subequations}}
\def\es{\end{subequations}}
\newcommand{\een}{\end{subequations}}
\newcommand{\ben}{\begin{subequations}}
\newcommand{\beq}{\begin{eqalignno}}
\newcommand{\eeq}{\end{eqalignno}}

\newcommand{\apj}{ApJ}
\newcommand{\mnras}{MNRAS}
\newcommand{\prd}{PhysRevD}
\newcommand{\jcap}{JCAP}
\newcommand{\aj}{AJ}
\newcommand{\apjs}{ApJS}

\newcommand\fverb{\setbox\pippobox=\hbox\bgroup\verb}
\newcommand\fverbdo{\egroup\medskip\noindent%
                        \fbox{\unhbox\pippobox}\ }
\newcommand\fverbit{\egroup\item[\fbox{\unhbox\pippobox}]}
\newbox\pippobox

\title{Testing $\mathbf{\Lambda}$CDM cosmology at turnaround:  \textit{where} to look for violations of the bound? }

\author[a]{D. Tanoglidis}
\author[a,b,c]{V. Pavlidou}
\author[a,b]{T.~N. Tomaras}
\affiliation[a]{Department of Physics, University of Crete, University of Crete, 71003 Heraklion,Greece} 
\affiliation[b]{Institute for Theoretical $\&$ Computational Physics, University of Crete, 71003 Heraklion,Greece}
\affiliation[c]{Foundation for Research and Technology - Hellas, IESL, Voutes, 7110 Heraklion, Greece} 
\emailAdd{dtanogl@physics.uoc.gr}
\emailAdd{pavlidou@physics.uoc.gr}
\emailAdd{tomaras@physics.uoc.gr}

\abstract{In $\Lambda$CDM cosmology, structure formation is halted shortly after dark energy dominates the mass/energy budget of the Universe. A manifestation of this effect is that in such a cosmology the turnaround radius --the non-expanding mass shell furthest away from the center of a structure-- has an upper bound. Recently, a new, local, test for the existence of dark energy in the form of a cosmological constant was proposed based on this turnaround bound. Before designing an experiment that, through high-precision determination of masses and --independently-- turnaround radii, will challenge $\Lambda$CDM cosmology, we have to answer two important questions: First, \emph{when} turnaround-scale structures are predicted to be  close enough to their maximum size, so that a possible violation of the bound may be observable. Second, which is the best \emph{mass scale} to target for possible violations of the bound. These are the questions we address in the present work. Using the Press-Schechter formalism, we find that turnaround structures have in practice already stopped forming, and consequently, the turnaround radius of structures must be very close to the maximum value \emph{today}. We also find that the mass scale of $\sim 10^{13} M_\odot$ characterizes the turnaround structures that start to form in a statistically important number density today --and even at an infinite time in the future, since structure formation has almost stopped. This mass scale also separates turnaround structures with qualitative different cosmological evolution: smaller structures are no longer readjusting their mass distribution inside the turnaround scale, they asymptotically approach their ultimate abundance from higher values, and they are common enough to have, at some epoch, experienced major mergers with structures of comparable mass; larger structures exhibit the opposite behavior. We call this mass scale the \emph{transitional} mass scale and we argue that it is the optimal for the purpose outlined above. As a corollary result, we explain the different accretion behavior of small and larger structures observed in already conducted numerical simulations. }

\begin{document}
%\begin{flushright}
%\end{flushright}

\maketitle
\flushbottom

\section{Introduction}
\setcounter{equation}{0}

The existence of a cosmological constant component in the matter/energy content of the Universe is currently inferred through the measurement of its accelerated expansion (using distant supernovae type Ia as standard candles, \cite{Riess,Perlmutter,Schmidt}) and also by the fact that the Universe seems to be flat (from observations of the cosmic microwave background, e.g., \cite{WMAP,Planck} for the latest results), while its matter density seems to be less than the critical (using measurements of baryon acoustic oscillations (BAO's), e.g., \cite{Addison et al, Beth} and references therein). We can call those two dark energy indicators  global or universal. They allow us to infer its existence by looking at the largest scales of the Universe. The observed acceleration of the Universe, however, also has a tremendous effect on --relatively-- small scales: in a $\Lambda$CDM universe the process of structure formation cannot last forever, due to the ``anti-gravitational" effect of the cosmological constant. Indeed, since the Universe is nowadays dark energy-dominated ($\Omega_{\mbox{\scriptsize{$\Lambda,0$}}}\cong 0.73 $), structure formation should be almost finished \cite{Busha1,Busha2}. 

Taking this into account, it is natural to try to construct a \emph{local} cosmological test of $\Lambda$, based on the process of structure formation. We can get the necessary insight into this process through the spherical top-hat model \cite{Peebles,GunnGott,Gunn}. A spherical perturbation in a $\Lx$CDM universe evolves as an
independent sub-universe, obeying its own Friedmann equation. If the matter density inside the region is enough to overcome the ``anti-gravitational" effect of the dark energy density, the perturbation will reach a maximum size (the turnaround radius) and then it will start to collapse. Formally it will collapse to a singularity; in practice, however,  its final radius is finite and is given by the value predicted by the virial theorem (the virial radius).

Recently such a local test was proposed \cite{PavTom}. In that work it is demonstrated that, in the $\Lambda$CDM cosmological model, there is a maximum value of the turnaround radius for a structure of mass $m$, which is independent of cosmic epoch and equal to
\begin{equation} \label{eq:eq11}\
R_{\mbox{\scriptsize{ta,max}}}= \left(\frac{3Gm}{\Lambda c^2} \right)^{1/3},
\end{equation}
where $G$ is the Newton's gravitational constant and $c$ is the speed of light. The cosmological test clearly follows from the fact that any \emph{non-expanding} structure of mass $m$ cannot have radius that exceeds the radius predicted by \eqref{eq:eq11} in the $\Lambda$CDM model. 
  
 Studying structure formation and designing a cosmological test at the turnaround (rather than virial) scale has many advantages : From a theoretical point of view, turnaround is a much better defined point in the spherical collapse model. Furthermore, non-sphericities are smaller at turnaround and that makes spherical collapse model more accurate when studying turnaround rather than virialized structures. Also, numerical simulations (e.g.,\cite{Cuesta}), theoretical (e.g.,\cite{Zemp,Anderhalden}) and observational arguments (e.g.,\cite{Miller}) show that no special physical meaning can be assigned to the the virial radius, as calculated through density-threshold arguments. It's not, for example, the boundary between a region of zero mean radial velocity with a region where material is in-falling or out-falling from the central region. On the other hand, turnaround is observationally defined as the non-expanding shell furthest away from the center of a bound structure. It separates the region where the gravitational attraction of the central structure is dominant, from the region where matter follows the general expansion of the Universe. In this sense, it is a very unambiguous boundary of a structure.

Although robust, the test would not be very useful if today we are far from the predicted --in the standard  cosmological model-- end of turnaround-scale structure formation. If our cosmology is close to --but not exactly-- $\Lambda$CDM, manifestations of this difference as violations of the bound are more likely to be present at an epoch when exact $\Lambda$CDM cosmology predicts that structures have almost achieved their maximum turnaround radius.  Numerical simulations, \cite{Busha1,Busha2}, suggest that structure formation is already quenched at the present epoch, but in our work we conclusively show that this is indeed the case at  turnaround scales. 

Finally, we locate the optimum mass scale of structures for the proposed violation searches, by studying which mass scales are most likely to have achieved their maximum size at a given epoch, especially the present epoch. 

Our work, is based on the semi-analytical Press-Schechter formalism  for the cosmic structure mass function, adapted appropriately to use the turnaround definition of a structure. A brief description of the Press-Schechter mass function is given in the next section.

Our paper is organized as follows: In section \ref{sec2} we discuss
the Press-Schechter formalism and the associated mass function, as
well as a rigorous way to define the ultimate mass function in a
$\Omega_{\mbox{\scriptsize{m}}}+\Omega_{\mbox{\scriptsize{$\Lambda$}}}=1$
cosmology. In section \ref{sec3}, we use this formalism to study the
end of turnaround-scale structure formation in that cosmology:
identify qualitatively different behaviors of different mass scales as
they approach their ultimate abundance; estimate the mass scale that
separates qualitatively different behaviors. In section \ref{sec4} we apply our results to the design of a new, local
cosmological test. We finally give a general discussion in \ref{sec5}.

\section{The mass function of turnaround structures} \label{sec2}

\subsection{The Press-Schechter mass function} \label{sec21}

The mass function of cosmic structures
is the comoving number density of structures with masses in the range $\left[m,m+dm\right]$. 
The Press-Schechter formalism \cite{pressscechter,Bond et al} and its excursion set formalism extensions   \cite{LaceyCole,PeacockHeavens,Jedamzik,Bower,Zenter}, provide semi-analytic approximations to cosmic mass functions. The Press-Schechter mass function is given by:
\begin{equation} \label{eq:eqps}\
\displaystyle \frac{dn}{dm}(m,a)\,dm= \sqrt{\frac{2}{\pi}}\frac{\rho_{\mbox{\scriptsize{m,0}}}}{m^2}\frac{\tilde{\delta}_{\mbox{\scriptsize{0,c}}}(a)}{\sigma(m)}\left|\frac{d \, \ln {\sigma(m)}}{d\, \ln{m}} \right|  \exp \left[ - \frac{\tilde{\delta}_{\mbox{\scriptsize{0,c}}}^{\,2}(a)}{2 \sigma^2(m)} \right]\,dm,
\end{equation}
and it is a function of mass, $m$, and cosmic epoch $a$. Although based on the simplifying assumption of spherical symmetry, it is generally in good agreement with numerical simulations (e.g., \cite{LaceyCole2}). More complex excursion set extensions \cite{ShethMoTormen,ShethTormen}, taking account of non-sphericities, show an even better agreement. Since we are going to focus on turnaround scales (where deviations from spherical symmetry are smaller), we will use the simpler Press-Schechter recipe.
 
  $\tilde{\delta}_{\mbox{\scriptsize{0,c}}}(a) $ is the linearly extrapolated, to the present epoch, overdensity of a structure that collapses (or reaches turnaround in our case) at a cosmic epoch $a$,  $\sigma^2(m)$ is the mass variance and $\rho_{\mbox{\scriptsize{m,0}}}$ is the present-day mean matter density of the Universe.

 We can define a characteristic mass, $m^{\star}$, by equating $\sigma(m^{\star})=\tilde{\delta}_{\mbox{\scriptsize{0,c}}}(a)$.
For $m \ll m^{\star}$, $\frac{dn}{dm}(m,a) $ has a power-law form. For $m \gg m^{\star}$, the abundance of structures is exponentially suppressed. From the above we can conclude that $m^{\star}$ gives us the mass scale of structures that start to form at epoch $a$.

\subsection{The criterion for turnaround} \label{ssec221}

Consider a small, homogeneous, spherical density perturbation, in an otherwise homogeneous universe, with $ \Omega_{\mbox{\scriptsize{m}}}+ \Omega_{\mbox{\scriptsize{$\Lambda$}}}=1 $. The behavior of the scale factor of the perturbation, $a_{\mbox{\scriptsize{p}}}$, as a function of the scale factor of the Universe, $a$, is described by the equation: 
\begin{equation} \label{eq:sfsf}\
\left(\frac{da_{\mbox{\scriptsize{p}}}}{da}\right)^2=\frac{a}{a_{\mbox{\scriptsize{p}}}}\frac{\omega a_{\mbox{\scriptsize{p}}}^3-\kappa a_{\mbox{\scriptsize{p}}}+1}{\omega a^3+1},
\end{equation}
where $\omega= \Omega_{\mbox{\scriptsize{$\Lambda$,0}}}/\Omega_{\mbox{\scriptsize{m,0}}}=\Omega_{\mbox{\scriptsize{m,0}}}^{-1}-1$. The constant $\kappa$ (curvature constant) is a constant which characterizes the magnitude and the sign of the perturbation. 
There is a minimum value of $\kappa$ a perturbation must have to be able to turnaround and collapse (by demanding the r.h.s. of (\ref{eq:sfsf}) to have a real and positive root):
\begin{equation} \label{eq:kappamin}\
\kappa_{\mbox{\scriptsize{min,coll}}}=\frac{3\omega^{1/3}}{2^{2/3}}.
\end{equation}
We can translate this to a minimum critical (linearly extrapolated to the present epoch) overdensity for turnaround and collapse, using the relation between  $\tilde{\delta}_0$ and $\kappa$ \cite{eke}:
\begin{equation} \label{eq:kdeltarel}
\displaystyle \kappa=\frac{(2\omega)^{1/3}}{3A \left[(2 \omega)^{1/3} \right]} \tilde{\delta}_0,
\end{equation}
where 
\begin{equation} \label{eq:Ax}
A(x)=\frac{(x^3+2)^{1/2}}{x^{3/2}}  \int^x_0 \left( \frac{u}{u^3+2} \right)^{3/2} du,
\end{equation}
is proportional to the linear growth factor in a $\Lambda$CDM Universe.
Substituting the minimum $\kappa$ for turnaround and collapse, given by eq.    \eqref{eq:kappamin}, we get the critical (minimum) overdensity, linearly extrapolated to the present epoch, that a perturbation must have to be able to reach turnaround in infinite time:
\begin{equation} \label{eq:mindelta}
\displaystyle
\tilde{\delta}_{\mbox{\scriptsize{min,coll,0}}}=\frac{9}{2}\,A\left[(2\omega)^{1/3}\right].
\end{equation} 
Inserting this value to eq. \eqref{eq:eqps} we obtain the \emph{ultimate} mass function of turnaround structures.

The threshold value in overdensity, for turnaround  at any other cosmological epoch can be obtained from the following relation:
\begin{equation}
\tilde{\delta}_{\mbox{\scriptsize{0,c}}}(a)=\frac{3 A\left [\left(2 \omega \right)^{1/3}\right ]}{2^{1/3}} \frac{1+\mu_{\mbox{\scriptsize{c}}}(a)}{\left[\mu_{\mbox{\scriptsize{c}}}(a) \right]^{2/3}},
\end{equation}
where we use again eq. \eqref{eq:Ax} and the quantity $\mu_{\mbox{\scriptsize{c}}}(a)$ is obtained by solving the equation
\begin{equation}
a= \omega^{-1/3}\left[\sinh {\cal{V}}_1(1,\mu)\right]^{2/3},
\end{equation}
with:
\begin{equation}
{\cal{V}}_1 (r, \mu) \equiv \frac{3}{2} \int_0^r \frac{\sqrt{x} \, dx}{\sqrt{(1-x)(-x^2-x+\mu)}}.
\end{equation}
For the derivation of these results and a more complete discussion about the spherical collapse model in $\Lambda$CDM cosmology see \cite{pikaifi}; for the spherical collapse model in other dark energy cosmologies see e.g.,  \cite{Pace et al}.

\section{Turnaround structure formation: results} \label{sec3}

We can now  use the Press-Schechter formalism described in \ref{sec21} to study how, in the $\Lambda$CDM cosmological model, the formation of cosmic structures comes to an end, focusing on turnaround rather than virialized structures. This will answer the question of how far are we, today, from the end of turnaround structure formation. 
Using eq. \eqref{eq:mindelta} we can get the ultimate (at $a=\infty$) mass function of turnaround structures. We can also get the mass functions for turnaround structures at any cosmological epoch $a$. We will compare the ultimate mass function with these, as well as with the mass function of virialized structures today.  We work in a cosmology with $\Omega_{\mbox{\scriptsize{m,0}}}=0.27$ and $\Omega_{\mbox{\scriptsize{$\Lambda$,0}}}=0.73$. The values of physical constants and cosmological parameters, where needed, were obtained from \cite{PPB}.

\subsection{Are we far from the end? Comparison of present-day mass functions with the ultimate mass function} \label{comp1}

\begin{figure}
\centering
\includegraphics[scale=0.7, clip]{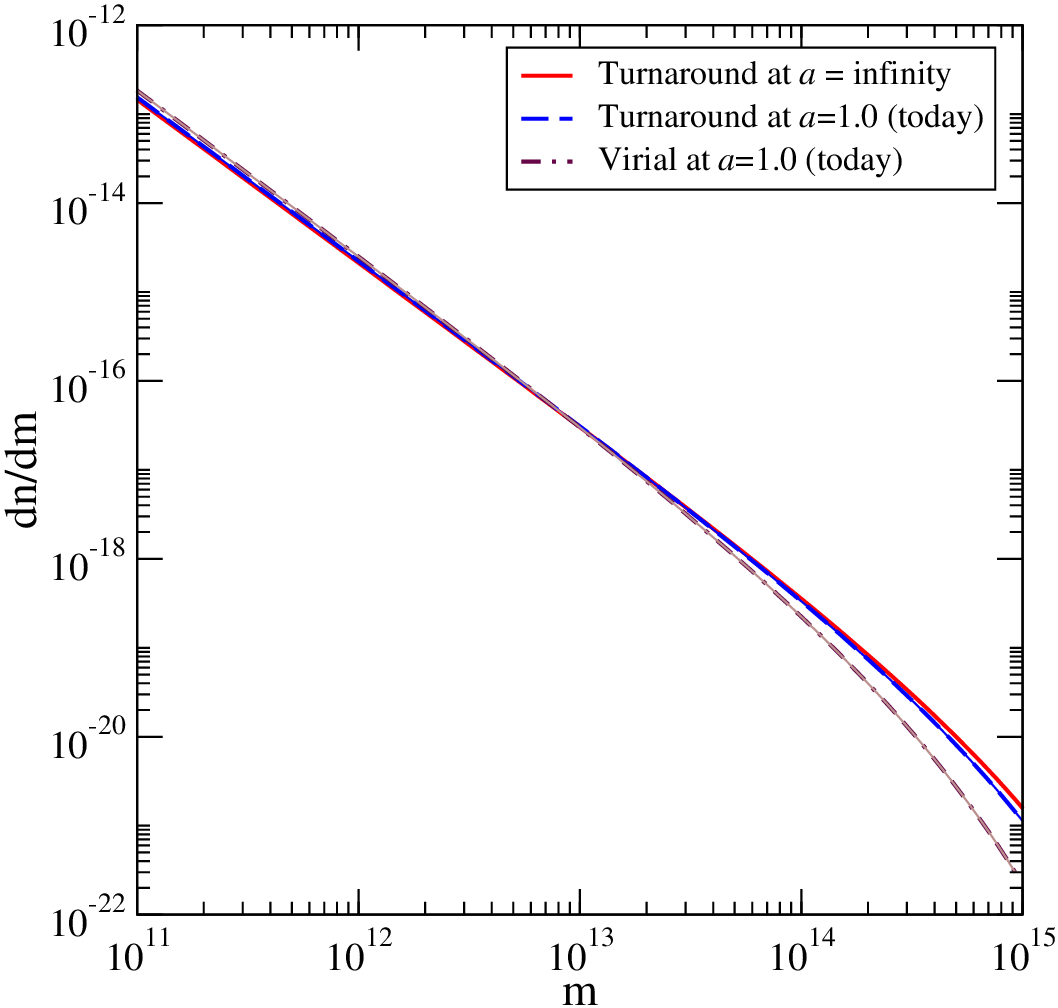} 
\caption{Press-Schechter mass functions; for turnaround structures (ultimate mass function and present-day mass function) and for virialized structures (present-day). The units of the mass functions ($dn/dm$ in the vertical axis) are $(M_\odot$ Mpc$^3)^{-1}$ and they are plotted as functions of mass, in units of solar masses, $M_\odot$.}
\label{fig:figure1}
\end{figure}

We start by comparing the ultimate mass function (for turnaround structures) with the  mass function for turnaround and virialized structures at the present cosmological epoch ($a=1$). We plot them together in Figure \ref{fig:figure1}. We have chosen to plot the mass functions in a mass range from $m=10^{11} M_\odot$ (the mass scale of a small galaxy) to $m=10^{15} M_\odot$ (the mass scale of a supercluster). The mass functions give  the comoving number density  per differential mass interval.

The two mass functions for turnaround structures are very similar. Especially in low-mass scales the two are almost identical and only in larger mass scales there is a small difference between them. Some deviation is seen after the start of the exponential suppression, which implies that structure formation is almost completed in the present cosmological epoch, if we define a structure by its turnaround overdensity. In the future, there will only be a small change in the number density of high mass structures, which is low anyway.

For comparison, we also present the mass function of virialized structures, for the present cosmological epoch. The divergence between that mass function and the ultimate mass function is greater than that of the two mass functions for turnaround structures. But, even in that case, the deviation is quite small and starts only for large mass scales, larger than $m_{\mbox{\scriptsize{div}}} \cong 2 \times 10^{13} M_\odot$, as can be seen in fig. \ref{fig:figure1}. 

From this behavior, we can conclude that structures today have almost reached their maximum turnaround radius. The comoving distribution of turnaround structures among various mass scales is almost the same as the final in the distant future. Structures with mass $> m_{\mbox{\scriptsize{div}}} \cong 5 \times 10^{13} M_\odot$ have not all virialized, despite having reached turnaround. Their remaining future evolution is to condense into denser virialized structures. We further quantify this qualitative behavior next.

\subsection{The path to the final number density of structures of various mass scales} \label{comp2}

In this section we investigate how different mass scales approach their final number density. To do so, we take, at some chosen mass scales, the ratio, denoted by \textit{r}, of the turnaround mass function at some cosmic epoch, $a$, over the ultimate turnaround mass function.
\begin{figure}
\centering
\includegraphics[scale=0.7, clip]{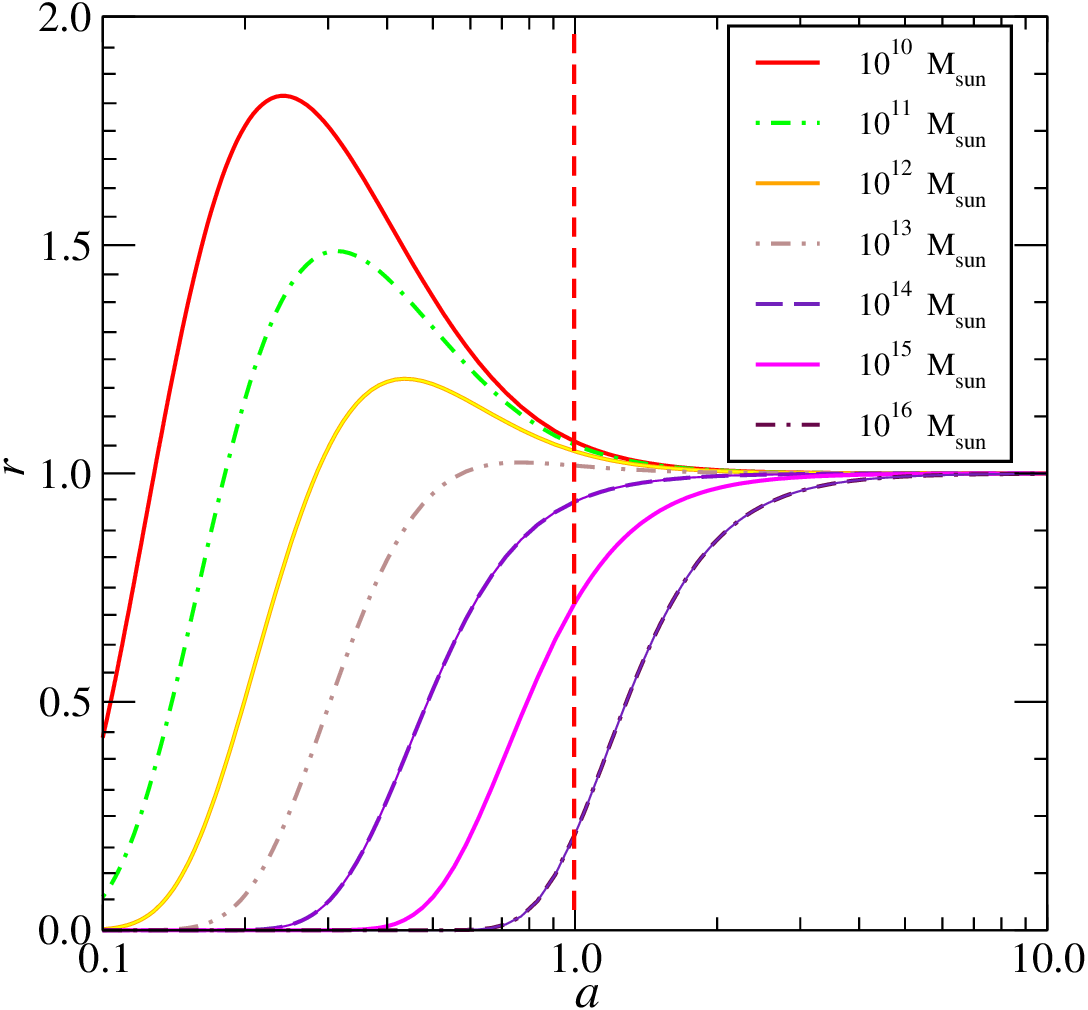} 
\caption{Evolution of the ratio, $r$, of the number density of turnaround structures  over the final number density, as a function of the cosmic epoch, labelled by the scale factor, $a$. We have plotted $r$ for some characteristic mass scales.}
\label{fig:figure2} 
\end{figure}
In Figure \ref{fig:figure2} we plot this ratio as a function of cosmic epoch, for mass scales starting from $10^{10} M_\odot$ and going up to $10^{16} M_\odot$ and for cosmic epochs starting from $a=0.1$ (in the past) and going up to $a=10.0$ (in the future). There are several interesting features to note on this plot:
\begin{itemize}
\item
Low-mass scales reach their final number density ($r=1.0$) at $a \sim 1$ (present cosmological epoch). These low-mass structures seem to reach their final number density at the same time (present) independently of their mass scale. Structures of mass $10^{10} M_{\odot}$ or $10^{12} M_{\odot}$, for example, have $r \cong 1.05$ and $r \cong 1.04$, respectively, at $a = 1.0$. All mass scales below $10^{14} M_{\odot}$ will readjust their abundance by less than $10 \%$ in the future.
\item 
For greater mass scales, there is a different trend. Very high-mass
structures reach $r=1.0$ later than high-mass structures. For example,
structures of $m=10^{16} M_{\odot}$ have a lower value of $r$ today
than structures with $m=10^{15} M_{\odot}$. These structures do not have a peak in the ``$r$ vs $a$" plot.
\item
High mass structures ($m \geq 10^{14} M_\odot$) always have lower number density than their final one. On the other hand, the number density of low-mass structures ($m \leq 10^{12}$) exhibits a different behavior: it starts lower than the final, it reaches a maximum, and then it decreases again, reaching its ultimate value  (corresponding to $r=1.0$) at  $a=\infty$, with $r<1.1$ today.
\end{itemize}

Why does the number density of low-mass structures decrease at late times while the
number density of high-mass structures increases? This is related to the hierarchical nature of structure formation.   Low mass structures enter the power-law
regime of the mass function before dark energy quenches structure formation. Consequently, they become, at some
point in cosmic history, relatively abundant. It is therefore likely
that they will encounter in their neighborhood, structures of similar
mass, and merge through major mergers to form larger structures. From
a Press-Schechter perspective, these structures are likely to be inside
overdensities of higher total mass that will eventually collapse. As
time progresses, those higher mass regions collapse to structures
(they reach the turnaround overdensity). Then, the initial low-mass
structures inside them are not counted any more as structures by the
Press-Schechter formalism. 
Conversely, high-mass structures are, and remain always, in the
exponentially suppressed part of the mass function. They are, and will
always be,  too rare to be nearby a similar or even bigger neighbour with significant probability. For this reason, they are not --again, statistically-- inside even larger overdensities that are going to be (turnaround) structures some day. 
In this interpretation, we expect the characteristic mass
scale separating the two regimes to lie near the mass scale where
exponential suppression sets in at the time when structure formation
comes to an end.

The existence of an ultimate mass function is a direct effect of dark
energy. Dark energy slows down --and finally halts-- structure
formation. This would not be the case, for example, in 
a flat, matter-dominated Universe, where structure formation continues
indefinitely: increasingly large mass scales collapse  and smaller
structures become part of them. It should be noted that in this case
an ``ultimate'' mass function cannot be defined in the Press-Schechter
formalism, as the smallest overdensity that will reach turnaround
approaches $0$ as $a$ grows without bound.

\begin{figure}
\centering
\includegraphics[scale=0.7, clip]{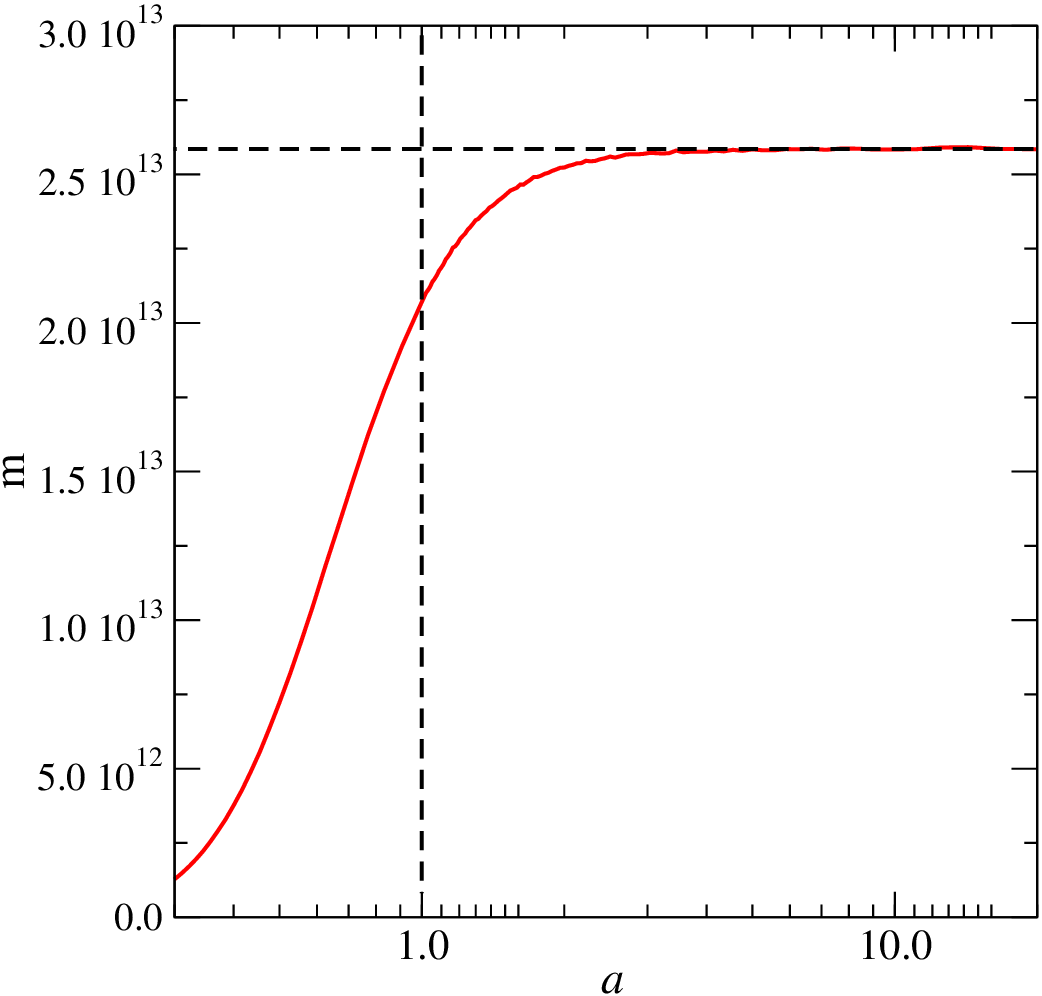} 
\caption{The mass scale, $m$, of the intersection of the Press-Schechter mass function at a cosmic epoch, $a$, with the ultimate Press-Schechter mass function. We have plotted this mass scale as a function of the cosmic epoch, $a$.}
\label{fig:figure3}
\end{figure}

It is interesting to identify the characteristic mass scale that
divides the two trends discussed above; i.e., the mass scale which
divides the ``low-mass" scales from the ``high-mass" scales. This will be
the first mass scale which does not exhibit a ``peak" such as the ones
that can be seen in Figure \ref{fig:figure2}. 
To find it, we plot in Figure \ref{fig:figure3} the mass of intersection of the --turnaround-- Press-Schechter mass function (at cosmic epoch $a$) with the ultimate mass function, as function of cosmic epoch, $a$. We can see that the mass of intersection approaches asymptotically characteristic value of
\begin{equation} \label{characteristic}
m_{\mbox{\scriptsize{char}}} \cong 2.6 \times 10^{13} M_\odot.
\end{equation}
Structures with mass greater than that of eq. (\ref{characteristic})  always have lower number density than their final one. 
At any cosmic epoch we expect lower-mass structures to be more abundant than ultimately (since some of them will become parts of greater structures), while we expect higher-mass structures  to be less abundant (bigger structures are going to be formed). At any epoch, there is a mass (the mass of the intersection of the mass function at that epoch, with the ultimate mass function) which divides the two trends. Following the evolution of this mass --which asymptotes to a constant value-- we find the mass scale of eq. (\ref{characteristic}).

\subsection{The transitional mass scale} \label{massscale}

In \S \ref{comp1} and \ref{comp2}, using the Press-Schechter formalism, we found differences in the behavior of low-mass and high-mass structures. Both $m_{\mbox{\scriptsize{div}}}$ and $m_{\mbox{\scriptsize{char}}}$, that divide structures according to their behavior, are of the order of $10^{13} M_\odot$. It seems that this mass scale is special. To make this statement even stronger, we calculate the characteristic mass  $m^{\star}$, described in sec. \ref{sec21}, for the current cosmological epoch (for turnaround and virialized structures) and at infinity (for turnaround structures). We have the following results:
\begin{equation}
m^{\star}_{\mbox{\scriptsize{0,turn}}} \cong 3.45 \times 10^{13} M_\odot,
\end{equation}
\begin{equation}
m^{\star}_{\mbox{\scriptsize{0,vir}}} \cong 1.68 \times 10^{13} M_\odot,
\end{equation}
and 
\begin{equation}
m^{\star}_{\mbox{\scriptsize{$\infty$,turn}}}  \cong 4.32 \times 10^{13} M_\odot,
\end{equation}
respectively. They correspond to the mass scales of structures that
start to form (having turnaround or becoming virialized) today and at
infinity. Once again, we observe that structure formation --as far as
we are considering turnaround scales-- has ended today, since
$m^{\star}_{\mbox{\scriptsize{0,turn}}}$ is very close to
$m^{\star}_{\mbox{\scriptsize{$\infty$,turn}}}$. Most interestingly,
and as we had anticipated when interpreting the qualitative
differences in the $r$ vs $a$ plot for the two classes of sources, all three  mass scales are of the order of $10^{13} M_\odot$.
Indeed, as far as our results based on the assumption of spherical symmetry are trustworthy, it seems that  the mass $m \sim 10^{13} M_{\odot}$ it is a special one in $\Lambda$CDM cosmology. Since this mass scale is the turning point between different behaviors of structures,  we call it the \emph{transitional} mass scale:
\begin{equation}
m_{\mbox{\scriptsize{transitional}}} \simeq 10^{13} M_\odot
\end{equation}
 we refer to structures with $m\lesssim m_{\mbox{\scriptsize{transitional}}} $ as low-mass and to  structures with $m\gtrsim m_{\mbox{\scriptsize{transitional}}}$ as high-mass structures.

\subsection{Manifestations of the transitional mass scale in simulations} \label{sims}

Our results offer a natural interpretation of effects already observed in numerical simulations.
Let us consider first low-mass structures. From Figures \ref{fig:figure1} and \ref{fig:figure2} we can see that they have almost reached, today, the maximum turnaround radius. Their number density is equal to the final. But also, from Figure \ref{fig:figure1}, all of them are also virialized structures. An important consequence of this is that low-mass structures  will not have a region of infalling material outside their dense, virialized cores. Just outside them, the Hubble flow will be the dominant feature. But for high-mass structures we expect a different behavior: the number density of turnaround structures is not equal to the number density of virialized structures. Statistically, most of high-mass structures  do have an infall region outside the central virialized region. Outside that infall region, the Hubble flow takes over, as before.

Such a behavior has already been observed in N-body simulations, \cite{Cuesta}. In that work, the authors sought the relation between the virial mass --as obtained from the standard overdensity threshold-- and a static mass, which is defined as the mass inside the innermost region with zero mean radial velocity, which better describes the dynamics of a virialized region. In doing so, they  explore the overall dynamical structure of dark matter halos, studying the behavior of the mean radial velocity as a function of the distance from the center of the halo. 
In the inner parts of halos (both low- and high-mass) the average radial velocities are zero. Similarly, at great distances the Hubble flow dominates. But the transition from the inner region to the Hubble flow is very different for low-mass and high-mass structures. For low-mass structures, the radial velocity is monotonically increasing as we are moving from the virialized region to the Hubble flow. But for high-mass structures, the situation is different, since outside the virialized region there is a region with large negative velocities, i.e. infall.

To  have an infall region outside the virial radius and before the turnaround radius is something expected when the process of structure formation is ongoing. The opposite behavior needs an explanation. Our findings suggest that this different behavior is a natural consequence of the halting of structure formation and growth, caused by the cosmological constant. Indeed the transition lies at the characteristic, transitional mass scale of about $10^{13} M_\odot$, exactly as expected by our discussion of the different behavior of low-mass and high-mass structures when the end of structure formation is concerned.

\section{Towards a local test of $\Lambda$CDM cosmology} \label{sec4}

Let us now discuss how our findings can be used to devise a test of our cosmological model, in the way proposed in \cite{PavTom}, by examining if there are structures that violate the bound given by eq. (\ref{eq:eq11}) --or a more refined bound, after taken into account the effects of non-sphericities, as also shown in that work.

We have demonstrated that, in the context of the accepted $\Lambda$CDM cosmological model, when defining structures by turnaround radius, structure formation has almost ended \emph{today}. This means that the turnaround radius cosmic structures have today, must be very close to the maximum predicted radius, if  $\Lambda$CDM is correct. Indeed, most nearby structures examined by \cite{PavTom} have (turnaround) radii very close to the maximum. This is evidence in favour of $\Lambda$CDM, in addition to the lack of any statistically significant violations of the bound in the observations studied by \cite{PavTom}.

Our results also allow us to address the following question: if a
survey were to be devised, dedicated to measuring independently masses
and turnaround radii of structures in the local universe, what would
be the best mass scale to target, in order to maximize the potential
of identifying structures possibly violating the $\Lambda$CDM turnaround bound
(eq. (\ref{eq:eq11}))?  We will argue that the most natural mass
scale to focus on is $m \sim 10^{13} M_{\odot}$. The reason is the
following: 

Statistically, in a bottom-up structure formation scenario, very small
structures start their evolution at higher overdensity values, and for
this reason are on average formed first. So at a given cosmic epoch,
say today, it will be the smallest structures that are the most likely
to have reached their maximum size. In Press-Schechter terms, lowering
the overdensity threshold will not, for these small structures,
incorporate regions with significant additional mass. If these
structures were close enough to big neighbours to be destined to be
``consumed'' by them, this is likely to have happened already. 

 However, focusing a survey that seeks to determine independently
 masses and turnaround radii on very small mass scales is
 impractical. The reason has to do with the determination of the
 turnaround radius. If structures are large enough to contain multiple
distinct observable substructures, such as  galaxies, then the
turnaround radius can be straight-forwardly determined: it is the
maximum radius that does not recede from the structure's
center. However, in smaller structures (a single small galaxy, for
example), it is not obvious how the turnaround radius could be
observationally identified, especially since these structures are
expected to be already dense, virialized structures. We are therefore looking for the {\em
  largest} mass scale at which, statistically, the maximum size has
been reached. From our discussion above, this mass scale is the mass
scale dividing the two qualitatively different behaviors: $m \sim
10^{13} M_{\odot}$ (groups of galaxies), or the transitional mass scale. Precision observations of
structures of this scale could thus allow us to test the existence of the cosmological constant \emph{locally} --using cosmic structures-- for the first time.

\section{Discussion} \label{sec5}

We have used the Press-Schechter formalism to study the end of structure formation in $\Lambda$CDM cosmology. Any cosmological model makes clear predictions about that era. For example, in a cosmology without a cosmological constant structure formation never comes to an end; structures are always formed, of greater and greater mass. On the other hand, the greater the value of the cosmological constant, the earlier structure formation stops. Similar predictions can be made \cite{PavTetrTom} for a dark energy equation of state different from  $w=-1$ or for the  quintessence model. We have chosen to concentrate on the simpler, currently consistent with data $\Lambda$CDM model. 

Due to the importance of the turnaround radius as a clear boundary of a structure, and also because it can be used to test the cosmological model, we have studied the end of structure formation concentrating on turnaround rather than virialized structures. We have found that the present-day turnaround mass function is very close to the ultimate mass function, which --\emph{inter alia}-- means that today, the turnaround radius of structures is very close to the maximum value predicted by \cite{PavTom}, eq. \eqref{eq:eq11}.

As a consequence of $\Lambda$, low- and high-mass structures exhibit different typical behaviors as cosmic structure formation draws to an end. The number density of low-mass and high-mass structures will approach the final from above and from below, respectively. This is because low-mass structures are most likely to have reached their maximum size and relaxed inside their virial size; high-mass structures are more likely to still be evolving -- either growing turnaround mass, or redistributing that mass within their turnaround radius until they reach virialization. As a result, low-mass structures will not have a region of infalling material around a denser core while structures with greater mass will exhibit such a behavior. The mass scale that separates low-mass structures and high mass structures is that of $10^{13} M_{\odot}$, and it is probably the best mass scale to search for potential violations of the upper bound in the turnaround radius. 

We have also found a new aspect of the ``cosmic coincidence" that dark energy is becoming dominant \emph{today}. There is one more manifestation of the fact that we  appear to be living in a very special time in the cosmic history. Structure formation has almost ended today; it is expected to be completely halted in the near future. The turnaround mass function today is the same as that of the distant future for most mass scales. 
However, as we have noted, the Press-Schechter formalism assumes spherical symmetry during the process of structure formation. For example, the overdensity value which defines a turnaround or virialized structure is obtained using the spherical collapse model, see \S \ref{ssec221}. This model, despite being a simplified approximation of the true process, gives insight on the process of structure formation. So the question arises: to what extend does the assumption of spherical symmetry affect our results? 

The effect of non-sphericities in the scales of interest has been studied analytically in \cite{Barrow} and numerically in \cite{Busha1} (see also the relevant discussion in \cite{PavTom}). Especially, in \cite{Busha1} a direct comparison between analytical predictions based on the spherical collapse model and results from numerical simulations is made. A main result is that for typical structures, with masses between $10^{14} M_\odot$ and $10^{16} M_\odot$, the sphere of gravitational influence is less than 30 $\%$ larger than the value predicted assuming spherical symmetry. Non-sphericities are also expected to be less important at turnaround scales --the scales we are mainly interested in-- compared to scales that characterize virialized structures. As the turnaround radius is defined as the non-expanding shell furthest away from the center of a bound structure, a mass shell in that region has been affected less by shearing forces, produced from the inhomogeneous distribution of matter, than a mass shell that has already collapsed under the influence of gravitational forces from the central overdence region. Finally, in $\S$\ref{sims} we saw that our results --especially the different behavior of structures with $m< 10^{13} M_\odot$ than that of structures with $m> 10^{13} M_\odot$-- agree with results obtained from numerical simulations. This further supports that the effect of non-sphericities at the spatial scales of interest is small.

In conclusion, we are confident that we have correctly identified the order-of-magnitude of the mass scale that should be targeted in a local test of $\Lambda$CDM cosmology, based on the maximum value of the turnaround radius. However, it is clear that the effect of non-sphericities \emph{on the value of the bound itself as a function of mass} should be carefully studied using numerical simulations before such a test is implemented in practice.

\acknowledgments

 This work was supported in part by European Union's Seventh Framework Programme under grant agreements 
(FP7-REGPOT-2012-2013-1) no 316165,
PIF-GA-2011-300984, PCIG10-GA-2011-304001, PIRSES-GA-2012-316788, the EU program ``Thales'' MIS 375734  and was also co-financed by the European Union (European Social Fund, ESF) and Greek national funds under the  ``ARISTEIA'' Action. 

%\newpage

%\section*{Bibliography}

\end{document}